\documentclass[fleqn,12pt,twoside]{article}
\usepackage{espcrc1}

\usepackage{graphicx}

\newcommand{\bi}{\begin{itemize}}
\newcommand{\ei}{\end{itemize}}
\newcommand{\be}{\begin{equation}}
\newcommand{\ee}{\end{equation}}

\newcommand{\bea}{\begin{eqnarray}}
\newcommand{\eea}{\end{eqnarray}}
\newcommand{\beastar}{\begin{eqnarray*}}
\newcommand{\eeastar}{\end{eqnarray*}}

\newcommand{\eq}[1]{~(\ref{#1})}
\newcommand{\eqq}[2]{~(\ref{#1},\ref{#2})}

\title{Stimulated and spontaneous relaxation in glassy systems}

\author{F. Ritort\\
Faculty of Physics, University of Barcelona, Diagonal 647, 08028 Barcelona, Spain and\\
Department of Physics, University of California, Berkeley CA 94720, USA\\
{\tt E-Mail:ritort@ffn.ub.es}}

\begin{document}

\maketitle

\begin{abstract}
{\em Recent numerical simulations of a disordered
system~\cite{CriRit03b} have shown the existence of two different
relaxational processes (called stimulated and spontaneous)
characterizing the relaxation observed in structural glasses. The
existence of these two processes has been claimed to be at the roots
of the intermittency phenomenon observed in recent experiments.  Here
we consider a generic system put in contact with a bath at temperature
$T$ and characterized by an adiabatic slow relaxation (i.e. by a
negligible net heat flow from the system to the bath) in the aging
state. We focus on a simplified scenario (termed as partial
equilibration) characterized by the fact that $T=0$ (where only the
spontaneous process is observable) and whose microscopic stochastic
dynamics is ergodic when constrained to the constant energy
surface. Three different effective temperatures can be defined: a)
from the fluctuation-dissipation theorem (FDT), $T_{\rm eff}^{\rm
FDR}$, b) from a fluctuation theorem describing the statistical
distribution of heat exchange events between system and bath, $T_{\rm
eff}^{\rm FT}$ and c) from a set of observable-dependent
microcanonical relations in the aging state, $T_{\rm eff}^{\rm
MR}$. In a partial equilibration scenario we show how all three
temperatures coincide reinforcing the idea that a statistical (rather
than thermometric) definition of a non-equilibrium temperature is
physically meaningful in aging systems.  These results are explicitly
checked in a simple model system.}
\end{abstract}

\section{Introduction}
Efforts to extend well established concepts in equilibrium
thermodynamics to the non-equilibrium domain have repeatedly appeared
many times in the past in different contexts. An idea that has
attracted the attention of physicists for quite a long time is the
concept of a temperature applied to non-equilibrium
states~\cite{CasJou03}. In equilibrium, the notion of temperature can
be covered from two different perspectives. On the one hand, there is
the thermodynamic approach where temperature is defined as a parameter
that characterizes classes of systems in mutual thermal
equilibrium. The usefulness of this {\em thermometric} definition
relies on the validity of the zeroth law of thermodynamics.  On the
other hand, there is a statistical approach (ensemble theory) where
temperature can be defined from the properties of individual systems
without any reference to the mutual equilibrium property. The {\em
statistical} temperature is defined as the inverse of the energy
gradient of the configurational entropy measured over a constant
energy surface. The maximum entropy postulate relates the statistical
notion of the temperature to the thermometric one.  The statistical
and thermodynamic concepts look equivalent but they are not as the
latter requires a specific behavior of the different systems when put
in mutual contact. The thermodynamic definition of a temperature
represents a stronger condition than the statistical one.  

An interesting category of systems that has recently received
considerable attention are glassy systems in their aging regime. The
aging regime is a slow relaxational process characterized by the
extremely small net amount of heat delivered by the system to the bath
in contact. It has been suggested~\cite{CugKurPel97} that in aging
systems a thermometric definition of non-equilibrium temperature is
meaningful for a thermometer responding to low-enough frequencies. We
suspect that this definition is too strong and might be wise to
investigate a {\em low-level} statistical definition of a
non-equilibrium temperature.

The purpose of this paper is to show that a statistical definition of
non-equilibrium or effective temperature (rather than thermometric)
can be rescued for a particular category of aging systems
characterized by adiabatic slow relaxation. In this category of
systems relaxation is imposed to be ergodic when constrained to a
given energy shell. At $T=0$ this condition defines what we term as
{\em partial equilibration} scenario. For this class of systems three
different definitions of an effective temperature are possible: a)
from the fluctuation-dissipation theorem (FDT), $T_{\rm eff}^{\rm
FDR}$; b) from the fluctuation theorem applied to the statistical
distribution of heat exchange events between system and bath, $T_{\rm
eff}^{\rm FT}$; c) from a set of microcanonical relations in the aging
state, $T_{\rm eff}^{\rm MR}$. In a partial equilibration scenario we
show how all these temperatures coincide reinforcing the idea that a
statistical definition of a non-equilibrium temperature is physically
meaningful in aging systems. We check all statements and results in a
simple model of glassy system where the partial equilibration scenario
holds and explicit computations can be done.

\section{Adiabatic relaxation}
\label{adiabatic}

Let us consider a system that is prepared in a non-equilibrium state
by placing it in contact with a thermal bath at low temperature
(quenching protocol). The system will relax and
release heat to the bath in a process that can span from several
minutes to millions of years. Only when the system has reached
equilibrium the net heat current from the system to the bath
vanishes.  All along the paper we will refer to this relaxational
regime as the aging regime and the corresponding non-equilibrium state
as the aging state. The time after the quench will be
referred as the age of the system and will be denoted by one or two of
the variables $s$ and $t$ depending on whether one-time or two-time
quantities quantities are considered. We adopt the convention $t>s$.

Several definitions and quantities seem appropriate to put the
discussion in perspective~\footnote{For a detailed presentation of
several of these concepts see \cite{CriRit03}}. Let us consider a system
described by an energy function $E({\cal C})$ where ${\cal C}$ denotes a
generic configuration. The system is in contact with a thermal bath and
the microscopic stochastic dynamics is both ergodic and satisfies the
detailed balance property.  $P_t({\cal C})$ will denote the probability for
the system to be in the configuration ${\cal C}$ at time $t$. The
average value of an observable $A$ at time $t$ will be denoted by $A(t)$
and is given by,
\be
A(t)=\sum_{{\cal C}}A({\cal C})P_t({\cal C})~~~.
\label{ad1}
\ee
Often the time argument in $A(t)$ will be dropped off and we will simply
write $A$, with the clear understanding that in general it is a time-varying
quantity.  $P({\cal C'},t|{\cal C},s)$ denotes the conditional or
transition probability for the system to be at configuration ${\cal
C'}$ at time $t$ if it was at configuration ${\cal C'}$ at
time $s$. The autocorrelation ($C_A(t,s)$) and response ($R_A(t,s)$)
functions are defined by,
\be C_A(t,s)=\sum_{{\cal C},{\cal C'}}P({\cal C'},t|{\cal C},s)A({\cal
C})A({\cal C'})~~~~~~~;~~~~~~~R_A(t,s)=\frac{\delta \langle A(t)\rangle}{\delta h(s)}
\label{ad2}
\ee
Both $C_A$ and $R_A$ can be decomposed into a stationary (fast) and
aging (slow) parts.  
In the large $s$ regime, for many aging systems a quasi-FDT relates the aging parts of\eq{ad2}
in terms of the fluctuation-dissipation ratio (FDR),
\be
R_A^{\rm ag}(t,s)=\frac{1}{T_{\rm eff}^{{\rm FDR}}(s)}\frac{\partial C_A^{\rm ag}(t,s)}{\partial
s}\theta(t-s)~~~.
\label{ad2b}
\ee
$T_{\rm eff}^{{\rm FDR}}(s)$ is a time-dependent effective
 temperature that has been shown to have
 some of the properties of a thermodynamic
 temperature~\cite{CugKurPel97}. In a weak ergodicity breaking
 scenario $C_A(t,s)$ decays to zero for $t-s\to\infty$ in a typical
 time that we denote by $\tau_{\rm decorr}(s)$~\footnote{This could be
 defined either as the integral correlation time
 $\int_s^{\infty}C_A(t',s)dt'$ or the value of $t-s$ for which
 $C_A(t,s)$ decays to $1/e$ of its maximum value $C_A(s,s)$.}. In
 general, $\tau_{\rm decorr}(s)\propto s^{\alpha}$ with $\alpha$ a
 given exponent. In many cases $\alpha=1$ appears to be a very good
 approximation, usually termed as simple aging.

We define the aging regime as adiabatic if the
fraction of heat released from the system to the bath goes
asymptotically to zero for time differences of order of the
decorrelation time,
\be
\Biggl|\frac{E(s+\tau_{\rm decorr}(s))-E(s)}{E(s)}\Biggr|\to 0~~~~~{\rm
for}\,\, s\,\, {\rm large\,\, enough}
\label{ad4}
\ee
Of course this relation is meaningful only in the aging regime where
the energy $E(s)$ is still far from its equilibrium
value~\footnote{Deviations from the power law divergence $\tau_{\rm
decorr}(s)\propto s^{\alpha}$ may provide an explicit check whether
the system has not left the asymptotic regime and aging is not
interrupted (in this last case the decorrelation time starts to
saturate to its value at equilibrium).}. The relation between the rate
decay of the energy and the frequency domain where FDT violations are
observed has been already pointed out in \cite{CugDeaKur97}.

\section{Stimulated and spontaneous relaxation as the origin of intermittency.}
\label{ss}
For an adiabatic regime as described in\eq{ad4} two type of heat
exchange processes are predicted to be observable depending on the
timescale of the measurement. According to\eq{ad4}, and for times of
the order of $\tau_{\rm decorr}(s)$ or smaller, the net heat flow
transferred from the system to the bath is exceedingly small, yet heat
fluctuations can be as big as if the system were in equilibrium at the
bath temperature. In such case there is a continuous heat exchange
between the system and the bath and energy fluctuations
$<E^2(s)>-<E(s)>^2$ (measured over $\tau_{\rm decorr}(s)$) are of the
same order of the energy content $E(s)$ and determined by the heat
capacity of the system at the bath temperature.  Were one to measure
at age $s$ the probability distribution $P_s(Q)$ of heat exchanged $Q$
between the system and the bath along intervals of a given duration
$\tau\sim \tau_{\rm decorr}(s)$, a Gaussian distribution would be
found with zero mean (as no net heat is transferred from the system to
the bath) and a variance $\sigma^2(T)$ which is independent of the age
$s$ but dependent on the temperature of the bath. This feature is
ubiquitous in glassy matter (e.g. structural glasses quenched at
low-enough temperatures) where the net heat flow from the system to
the bath is unobservable, yet heat is quickly exchanged with the bath
as thermal conductivity is high~\footnote{The simplest evidence in
favor of this statement is a piece of silica vitrified at room
temperature (i.e. not yet equilibrated) whose temperature can be felt
by touching it with the hand.}. For all practical purposes the system
looks equilibrated at the temperature of the bath and a thermometer
put in contact with it would measure the bath temperature.  We call
this heat exchange process {\em stimulated} as it originates in the
existence of physical processes thermally excited by the bath.

\begin{figure}[htb]
\begin{center}
\includegraphics[scale=.39]{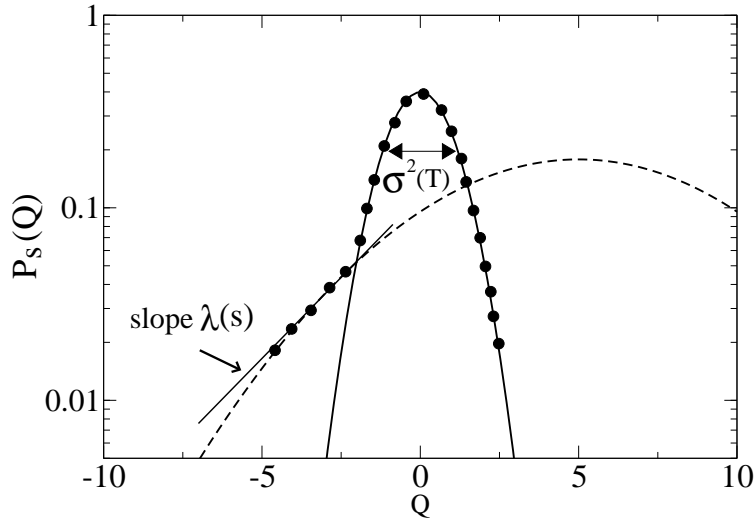}
\end{center}
\caption{\small The origin of intermittency: Two mechanisms of relaxation,
stimulated (continuous line) and spontaneous (dashed line) characterize 
the overall relaxation. $Q$ stands for the heat transferred from the
bath to the system (the sign indicating the direction of flow, if $Q<0$ heat flows from the system to the bath) along
regularly spaced time intervals and $P_s(Q)$ the corresponding
probability distribution at age $s$.  The black points illustrate
what experimental measured data would look like. In the simplest
scenario both processes are described by a Gaussian distribution. The
stimulated component has a variance $\sigma^2(T)$ dependent on the
bath temperature but independent on the age.  The spontaneous
component shows an exponential tail in the $Q<0$ side (shown as a
continuous line) whose slope $\lambda$ is age dependent. The
spontaneous process for $Q>0$ is unobservable as relaxational dynamics
is constrained by the existence of a net heat transfer from the system to the bath.}
\label{intermittency}
\end{figure}

For times much larger than $\tau_{\rm decorr}(s)$ the average net heat
flow from the system to the bath is not negligible, a clear
consequence that the system has not yet equilibrated. Most of the
exchange processes that occur along these timescales are stimulated by
the bath, however other exchange events do not fall into the previous
category and follow a completely different distribution. Contrarily to
the stimulated process, this distribution is expected to display an
exponential tail whose width $\lambda(s)$ depends on the age of the
system $s$. However this process is not excluded to be described by a
Gaussian distribution as well. Indeed an exponential tail is
characteristic of Gaussian distributions centered around a finite
value.  We call this new process {\em spontaneous} as its statistical
description is not determined by the temperature of the bath but
rather by the fact that the system has been prepared in a
non-equilibrium state.  Figure~\ref{intermittency} illustrates typical
distributions for the stimulated and spontaneous processes.  The
existence of two different kinds of heat exchange process (and
therefore two different heat exchange distributions) occurring along
widely separated timescales has been recently numerically verified in
the context of a simple spin-glass model~\cite{CriRit03b}. It has been
suggested that the existence of these two processes is at the roots of
the intermittency phenomenon observed in glasses~\cite{Cil03} and
colloids~\cite{Cip03}~\footnote{For the latter it might seem more appropriate to
speak about stress release rather than heat exchanged as compaction
(rather than heat dissipation) is the main relaxational process.}.

\section{Statistical (microcanonical) description of the aging state}
\label{micro}
Along the rest of the paper we will analyze in detail and idealized
particular case of the more general previous scenario. In doing this we aim
to understand fundamental issues behind a possible statistical, rather
thermodynamical, description of the glassy state.  We will
consider in detail a situation where the stimulated process does not
exist and only the spontaneous process is observed in the adiabatic
relaxation. We will refer to it as the {\em partial equilibration}
scenario. It is then possible to prove that effective temperatures,
with a precise statistical meaning, do emerge. This is the content
of the next Subsections, \ref{partial},\ref{ft},\ref{teff}. In
Sec.~\ref{om} we will focus our attention on the oscillator model (OM)
\cite{BonPadRit97} as a simple case where all results of the current
Section can be explicitly verified.

\subsection{Partial equilibration scenario}
\label{partial}
\begin{figure}[htb]
\begin{center}
\includegraphics[scale=.3]{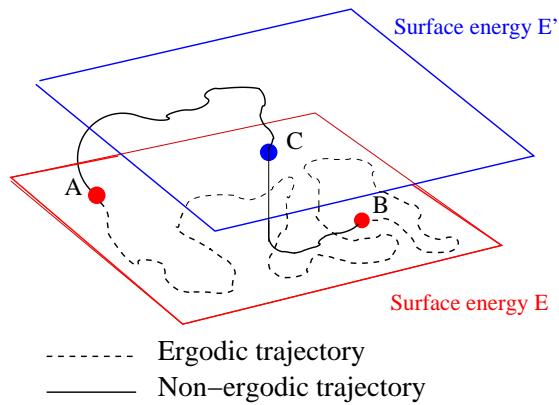}
\end{center}
\caption{\small Schematic picture of the type of ergodic dynamics considered
in a partial equilibration scenario. Any phase-space points $A,B$
contained in the surface of energy $E$ must be connected by a path
lying in that surface (dashed line).  Note that other paths
(continuous line going through $C$) that we call non-ergodic (as they change the energy
of the trajectory) might also exist as the original dynamics of the
system coupled to the bath is ergodic.}
\label{ergo}
\end{figure}

To suppress the stimulated process the bath is put at zero temperature
(therefore, no heat can flow from the bath to the system). However, to
keep the system in an adiabatic relaxational regime\eq{ad4} (avoiding
the aging regime to be quickly interrupted due to the presence of
forever-lived metastable configurations) we require an additional
condition: {\em Dynamics must be ergodic when the system is
constrained to move in any constant energy surface~\footnote{A more
precise statement requires dynamics to be ergodic within the energy
shell between $E$ and $E+\Delta E$ where $\Delta E$ is finite in the
large volume limit.}}. This condition is illustrated in
Fig.~\ref{ergo}. In this case, it is easy to prove that relaxation
cannot arrest at $T=0$ as it is always possible for the system to
decrease its energy by moving along the constant energy surface
until a favorable downhill move occurs. Models that fall into this
category have been termed as models with purely entropic barriers, the
OM~\cite{BonPadRit97} being a prominent example.
Eq.\eq{ad4} states that in an adiabatic regime the energy remains
practically constant during timescales of the order of the
decorrelation time $\tau_{\rm decorr}(s)$. Therefore, as correlations
decay in timescales much shorter than the energy does, the system can
explore a large number of configurations within a given region of
phase space.  In case only spontaneous relaxation takes place, the system
is constrained to reach a quasi-stationary state where all
configurations lying in a given constant energy surface have the same
probability. Equiprobability is guaranteed by the fact that
microscopic dynamics over the constant energy surface is ergodic and
satisfies detailed balance, i.e. it is reversible over the constant
energy surface. The
process where a given region of the constant energy surface is sampled
according to the microcanonical ensemble (i.e. all configurations
lying in that surface have the same probability to be visited) will be
termed {\em partial equilibration}.

For later purposes we define the complexity ${\cal S}_E(A)$ by the relation
\be
{\cal S}_E(A)=\log(\Omega_E(A))=\log\Bigl[\sum_{{\cal C}}\delta(E-E({\cal C}))\delta(A-A(\cal C))\Bigr]
\label{m1}
\ee
where $\Omega_E(A)$ stands for the number of configurations with
energy $E$ and observable $A$~\footnote{As we are counting
configurations, ${\cal S}_E(A)$ is nothing more but the usual
configurational entropy. However, we deliberately avoid to use this
term and prefer to talk about complexity. We have in mind the most
general case (not addressed in this paper, see
instead~\cite{CriRit03b}) where partial
equilibration is established among regions or components of phase
space. For this case the use of the term {\em configurational entropy}
could be misleading.}. For the total complexity $S(E)$ we have,
\be
S(E)=\log(\Omega(E))=\log\Bigl[\sum_{{\cal C}}\delta(E-E({\cal C}))\Bigr]
\label{m2}
\ee
where now $\Omega(E)$ is the number of configurations with energy $E$
irrespective of the value of the observable $A$, $\Omega(E)=\int dA
\Omega_E(A)$. In\eq{m1} we use the energy label $E$ as a subindex and
not as an additional argument, just to stress its key role in the dynamics 
compared to that of other observables.  Note that we could equally
well use the subindex $s$, as for a given age $s$ the average energy of
the system is fully determined by the dynamics.

\subsection{A fluctuation theorem in the aging state}
\label{ft}
In a partial equilibration scenario transitions between configurations
lying at different constant energy surfaces (i.e. transitions that
increase or decrease the energy) are not equiprobable. However, after
applying an external perturbation, a
shift of energy levels takes place and some configurations, initially
belonging to different energy surfaces, may end up into the same
one. We show below in Sec.~\ref{teff} how the response of the system to an external
perturbation is determined by the density of energy levels just around
the reference value $E(s)$. Accordingly, transition rates between
configurations having different energies $E,E'$, were they
equiprobable, are described by the following microcanonical relation,
\be
\frac{P(\Delta E)}{P(-\Delta E)}=\frac{\Omega(E')}{\Omega(E)}\label{n1}
\ee
where $\Delta E=E'-E$ is the (intensive) energy difference between
both configurations.  It is important to point out that $P(\Delta E)$ is
not the actual transition rate, but the rate the
system would display if transitions between configurations having
different energies were equiprobable, a situation that is encountered
only after shifting the energy levels (by applying an external
perturbation) and redistributing them into a unique energy surface.

Eq.\eq{n1} has the form of a FT since it describes the ratio of
rates in the forward $E\to E'$ and the
reverse $E'\to E$ directions. Fluctuation theorems, similar to\eq{n1}, have
been derived in other contexts, for instance in systems in steady
states~\cite{EvaSea02} or systems arbitrarily perturbed from an initial
equilibrium state~\cite{Crooks98}. From\eq{m2} we can write,
\be
\frac{P(\Delta E)}{P(-\Delta E)}=\exp\Bigl[\Bigl(\frac{\partial
S(E)}{\partial E}\Bigr)\Delta E)\Bigr]=\exp\Bigl[\frac{\Delta E}{T_{\rm
eff}^{\rm FT}(s)}\Bigr]
\label{n2}
\ee
where we have expanded the complexity $S(E)$ and kept only the first
term in $\Delta E$~\footnote{As $\Delta E$ is an intensive quantity,
higher order powers in $\Delta E$ should be included
in\eq{n2}. However they are not relevant for what is addressed in the
present paper. A similar situation is later encountered in\eq{n2}. See
the ensuing footnote (\ref{higherorderterms}) for a more detailed
explanation.}. The factor $T_{\rm eff}^{\rm FT}(s)$ in the exponent in
the r.h.s of\eq{g2} defines an effective temperature,
\be
\frac{1}{T_{\rm eff}^{\rm FT}(s)}=\Bigl(\frac{S(E)}{\partial E}\Bigr)_{E=E(s)}
\label{n3}
\ee
where we specifically include the subindex $s$ to denote its time
dependence through the value of the energy $E$ at age $s$. The
super-index FT indicates that this effective temperature is derived from
the fluctuation theorem\eq{n1}. For a Gaussian $P(\Delta E)$ the
value of $T_{\rm eff}^{\rm FT}(s)$ can be shown to be proportional to
the width of the exponential tail $\lambda(s)$ as depicted in
Fig.~\ref{intermittency}. This connection has been exploited in
\cite{CriRit03b} as a possible way to estimate the effective temperature
from intermittency measurements.

Eqs.\eqq{n1}{n2} may look as simple detailed balance but it is
not. Detailed balance is a inherent property of the microscopic
dynamics. Two are the main differences between\eqq{n2}{n3} and
microscopic detailed balance. In the former we assumed equiprobability
between transitions with identical energies and this is not guaranteed
if not in the partial equilibration scenario. Moreover the factor\eq{n3} in the exponent of\eq{n2} is not the temperature of the bath as
implied by detailed balance, but a quantity that is solely determined by the
dynamics.
 
We contend to show that $T_{\rm eff}^{\rm FT}(s)$ as derived from the
fluctuation theorem coincides with the effective temperature derived
from the FDT relations that link correlations and responses in the aging
regime. The origin of this connection has been already
mentioned. Eq.\eq{n1} links transitions between configurations with
different energies. These transitions can be probed only by lifting the
energy of the different configurations after applying an external field.

\subsection{Microcanonical relations and effective temperatures in the aging state.}

\label{teff}

Let us consider $A$ to be any observable of the system that can take
different values in a given surface of constant energy $E$.  The
equiprobability assumption implies that the transition rates between
configurations with different observable values $A,A'$, at a given age
$s$ (when the surface explored has energy $E$), satisfy the
following relation,
\be
\frac{W_E(\Delta A)}{W_E(-\Delta A)}=\frac{\Omega_E(A')}{\Omega_E(A)}
\label{g1}
\ee
where $\Delta A=A'-A$ and $\Omega_E(A)$ was defined in\eq{m1}. The ratio of rates\eq{g1} is
therefore age dependent as the value of $E$ changes with the age of the
system. 
Eq.\eq{g1} says that the rate for the observable $A$ to change its
value in $\Delta A$ within the surface of constant energy $E$, when
going from the value $A$ to the value $A'$, is proportional to the
number of configurations with value $A'$ at the energy
$E$. Although\eq{g1} describes rates there is no explicit reference to
any timescale. In fact, we use the term $W$ for these rates (as compared
with the term $P(\Delta E)$ used in\eq{n1}) to stress the fact that these
are rates rather than probabilities (i.e. they have dimensions of
a frequency). Note that the difference between using $P$ or $W$ is minor
as any timescale drops off when considering the ratio among the forward
and reverse rates. Again, as was done for\eq{n1}, 
we expand the term in the r.h.s. of\eq{g1} around $A'=A$ by using
\eq{m1} and consider only the first term in the expansion,
\be
\frac{W_E(\Delta A)}{W_E(-\Delta A)}=\exp\Bigl[\Bigl(\frac{\partial
{\cal S}_E(A)}{\partial A}\Bigr)\Delta A +{\cal O}((\Delta A)^2)\Bigr]
\label{g2}
\ee
The variation $\Delta A=A'-A$ is intensive, therefore all powers of
$\Delta A$ enter in\eq{g2} even in the thermodynamic limit. However, as
we will see later, only the first term is relevant for the
emergence of effective temperatures. Here we will not discuss the
possible relevance of higher order terms~\footnote{\label{higherorderterms}Higher order terms in $\Delta A$ are expected to contain information about
the stability of the aging state and thermally induced fluctuations of
the effective temperature.  Contrarily to the temperature of the
bath (which cannot fluctuate) the latter might display
fluctuations even in the large $V$ limit.}.
Relation\eq{g2} says that dynamics evolves towards configurations that
have a higher value of the complexity ${\cal S}_E(A)$, i.e. towards the
maximum value $A^*(E)$ of ${\cal S}_E(A)$,
\be
\Bigl(\frac{\partial{\cal S}_E(A) }{\partial A}\Bigr)_{A=A^*(E)}=0
\label{g3}
\ee
As the value of $E$ is time dependent it is also the value of the
maximum $A^*(E)$.  In a partial equilibration scenario, $A$ relaxes
(along timescales of order of $\tau_{\rm decorr}(s)$) to the value
$A^*(E)$ given in\eq{g3}. This value changes in time as the energy
decreases and is slaved to the time evolution of the energy. As the
energy decreases much slower than $A$ relaxes to $A^*(E)$ (c.f.\eq{ad4}),
reversibility holds in the aging regime,
\be
W_E(\Delta A)=W_E(-\Delta A)~~~~.
\label{g4}
\ee
An interesting and special class of observables are those called {\em
neutral} where, for large enough times, the value of $A^*(E)$ is
independent of $E$ (i.e. of the age of the system) $A^*(E)=A^*$.
Therefore the value $A^*$ must coincide with the equilibrium value $A^{\rm
eq}$ if the system has to equilibrate. Neutral
observables have the interesting property that they reach a stationary
value exponentially fast and their time evolution is not slaved to
that of the energy.  The most prominent example of neutral observables
is the wave vector density $\rho_k(t)$ in supercooled liquids which
stays negligible at all times as there is no long range order that
develops in the amorphous glass state. The dynamics of a neutral
observable is therefore quite easy to visualize. Starting from any
initial configuration the dynamics of the system quickly evolves
towards $A^*$ and stays there forever.

Non-neutral observables are expected to relax fast to their value
$A^*(E)$ and display an interesting non-monotonic behavior as the value
$A^*(E)$ changes with the age $s$. For instance, if the system starts
far from equilibrium but with a value of $A$ corresponding to the
equilibrium value $A^{\rm eq}$, initially the system will depart from
this value, and follow the time evolution of $A^*(E)$ to eventually come
back again to $A^{\rm eq}$ after equilibrating. This effect has been
observed in structural glasses where the volume density displays
non-monotonic behavior in the glass state and is known as the Kovacs
effect~\footnote{This effect has been studied in different models of
glasses, for instance~\cite{BerBouDroGod03,Buhot03}}.
\begin{figure}[htb]
\includegraphics[scale=.3]{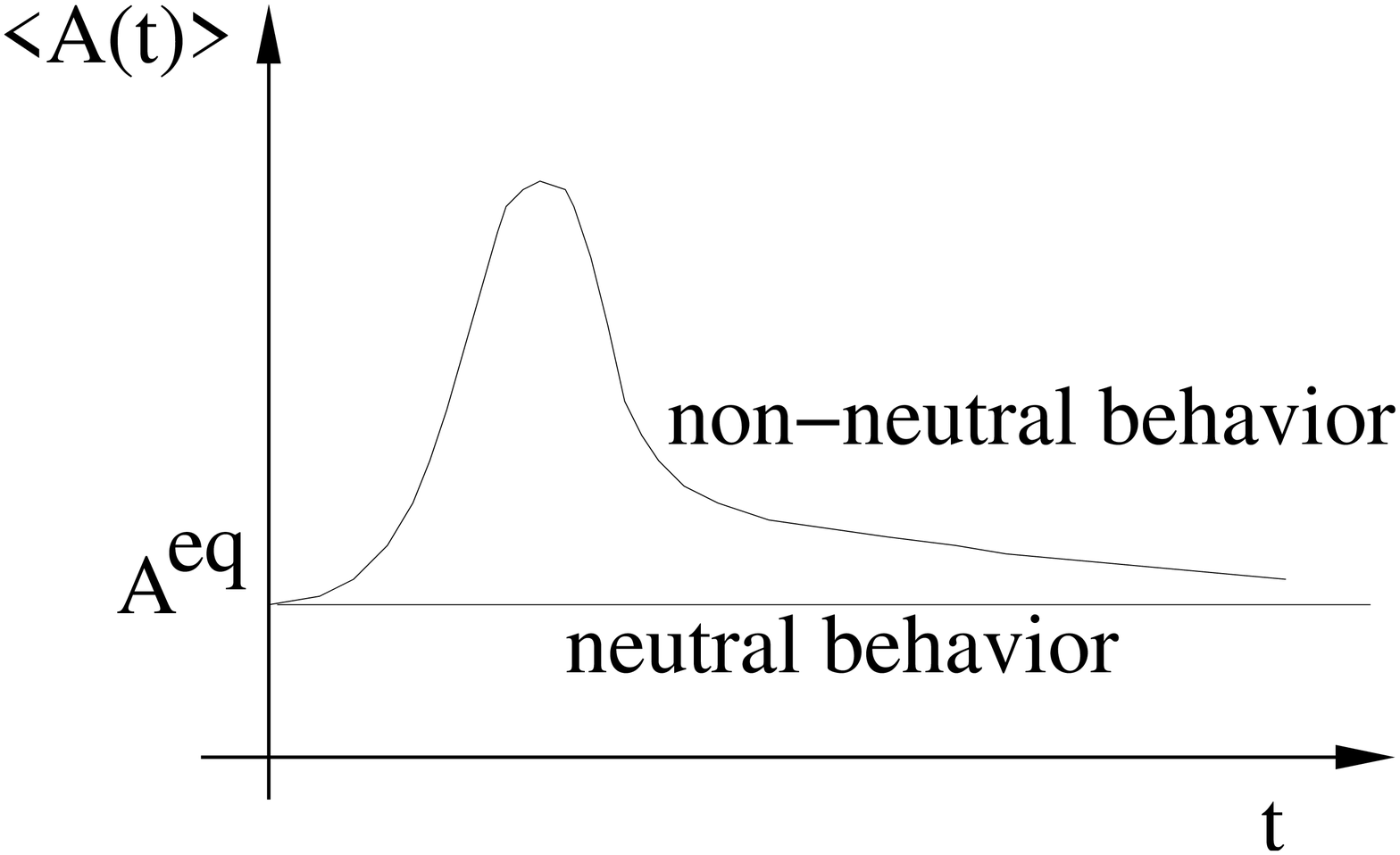}
\caption{\small Time evolution of neutral and non-neutral
observables. Explanation is given in the text.}
\label{neutral}
\end{figure}

We are now in a position to understand the emergence of effective
temperatures as usually derived from fluctuation-dissipation relations. 
Suppose now that at time $s$ an external field of intensity $h$ coupled
to the observable $A$ is applied to the system. In the presence of a
field the energy of a configuration ${\cal C}$ is shifted by the Zeeman term,
\be
E({\cal C})\to E({\cal C})-hA({\cal C})~~~~~.
\label{g5}
\ee
Under the field, the surface of constant energy $E$ does not
coincide anymore with that at zero field, meaning that configurations with
identical energy at zero field (i.e. lying in a given surface of
energy $E$) get shifted by different amounts in a field (i.e. finish
into different surfaces with different values of $E$). 
In particular, for a given value of the energy $E$, from all configurations
initially lying in the surface at $h=0$, after switching the field
some configurations leave the surface, others come onto the surface,
finally others stay there (i.e. those with $A({\cal
C})=0$).

According to\eq{g5}, just after the field has been switched on,
configurations with a larger value of $A$ decrease their energy relative
to their energy at $h=0$ and configurations with a lower value of $A$
increase their energy.  However, the complexity $S(E)$\eq{m2} is a monotonic
function of the energy, therefore configurations with higher energy are more
numerous than those with lower energy. As a result of the action of the
field, the number of configurations $\Omega_{E}^h(A)$ with a given value
of $A$, that lie inside the surface of constant energy $E$,
monotonically increases with $A$,
\bea
\Omega_{E}^h(A)=\int dE' \Omega_{E'}(A)
\delta(E-E'+hA)=\Omega_{E+hA}(A)=\nonumber\\
\Omega_E(A)\Bigl[1+\frac{\partial \log(\Omega_E(A))}{\partial E}Ah+{\cal
O}(h^2))\Bigr]=
\Omega_E(A)\exp\Bigl[\Bigl(\frac{\partial{\cal S}_E(A)}{\partial E}\Bigr)Ah+{\cal
O}(h^2)  \Bigr]~~~.
\label{g6}
\eea
Using the equivalent of\eq{m1}, ${\cal S}_E^h(A)=\log(\Omega_{E}^h(A))$,
we get,
\be
{\cal S}_E^h(A)={\cal S}_E(A)+\Bigl(\frac{\partial{\cal S}_E(A)}{\partial E}\Bigr)Ah+{\cal
O}(h^2)={\cal S}_E(A)+\beta_{\rm eff}(E,A)Ah+{\cal
O}(h^2) 
\label{g7}
\ee
where we have defined,
\be
\beta_{\rm eff}(E,A)=\Bigl(\frac{\partial{\cal S}_E(A)}{\partial E}\Bigr)_{E=E(s),A=A(s)}
\label{g8}
\ee
As ${\cal S}_E(A)$ is a monotonic increasing function of $E$,\eq{g6}
indicates that relaxation in a field is pushed towards
configurations with progressively increasing values of $A$.
In a field partial equilibration occurs
again in the new surface of energy $E$ as all configurations there
contained remain equiprobable. The new \eq{g1} reads,
\be
\frac{W_E^h(\Delta A)}{W_E^h(-\Delta
A)}=\frac{\Omega_E^h(A')}{\Omega_E^h(A)}=\frac{W_E(\Delta
A)}{W_E(-\Delta A)}\exp(\beta_{\rm eff}(E,A)\Delta A \,h)
\label{g10}
\ee
where\eqq{g6}{g7} have been used. A word of caution in the derivation of
\eq{g10} is in place. We have considered $\beta_{\rm
eff}(E,A)=\beta_{\rm eff}(E,A')$. This is justified as the quantity
${\cal S}_E(A)$ entering into the definition\eq{g8} is {\em extensive},
i.e. proportional to the volume $V$ of the system, while $\Delta A=A'-A$ is
an intensive quantity. In the large $V$ limit, the partial derivative
\eq{g8} is the same whether it is taken at $A$ or $A'$ as the difference
$\Delta A=A'-A$ is intensive so $\beta_{\rm eff}(E,A)-\beta_{\rm
eff}(E,A')\sim {\cal O}(1/V)$.

Eq.\eq{g10} is reminiscent of detailed balance, however the same
remark has to be made here as was done in Sec.~\ref{ft}.  The quantity
$\beta_{\rm eff}(E,A)$ is not the temperature of the bath anymore but
a time dependent value as $E(s)$ and $A(s)$ change in
time.
The quantity  $\beta_{\rm eff}(E,A)$ defines an effective
temperature as obtained from the microcanonical relation (MR)~\eq{g10},
\be
\frac{1}{T_{\rm eff}^{{\rm MR},A}(s)}=\beta_{\rm eff}(E,A)=\Bigl(\frac{\partial{\cal S}_E(A)}{\partial E}\Bigr)_{E=E(s),A=A(s)}~~~.
\label{g11}
\ee
As $A(s)=A^*(E(s))$ (see\eq{g3} and the ensuing discussion) we can
replace the complexity ${\cal S}_E(A)$ by the total complexity $S(E)$ in the
r.h.s of\eq{g11}. Notably this gives a value of $T_{\rm eff}^{{\rm MR}}(s)$
that is independent on the type of observable so the label $A$ in the l.h.s of\eq{g11} drops off,
\be \frac{1}{T_{\rm eff}^{{\rm MR}}(s)}=\beta_{\rm eff}^{{\rm
MR}}(s)=\Bigl(\frac{\partial{\cal S}_E(A))}{\partial
E}\Bigr)_{E=E(s),A=A(s)}=\Bigl(\frac{\partial S(E)}{\partial
E}\Bigr)_{E=E(s)}~~~,
\label{g13}
\ee
which coincides with the result obtained from the FT\eq{n3}.

From relation\eq{g10} it is now possible to link  the 
autocorrelation and response functions\eq{ad2} through via the FDR, where the effective
temperature\eq{g13} appears explicitly,
\be
R_A(t,s)=\frac{1}{T_{\rm eff}^{{\rm MR}}(s)}\frac{\partial C_A(t,s)}{\partial s}\theta(t-s)~~~.
\label{g12}
\ee
We are not going to show here the details of this derivation but only
mention the main steps. The proof follows the scheme of the derivation
shown in Section~3 of Ref.~\cite{CriRit03} for the standard derivation
of FDT where\eq{g10} here corresponds to (49) there. In
\cite{CriRit03} it is shown how the response function can be
decomposed in two different contributions called $R^{(1)}$ and
$R^{(2)}$. In equilibrium, when FDT holds, $R^{(1)}$ vanishes and
$R^{(2)}$ gives the equilibrium response.  In the partial
equilibration scenario,\eq{g12} is obtained whenever $R^{(1)}$
vanishes. Close inspection of (53) in \cite{CriRit03} shows that this
term vanishes only if the perturbing field does not modify, to linear
order in $h$, the trapping time distribution measured over transitions
starting at the surface of energy $E$.  More precisely, consider a
sample of all trapping times for configurations of age equal to
$s$. As dynamics is stochastic, the same quenching protocol will
generate different configurations with values $E,A$ in the vicinity of
their dynamical averages $E(s),A(s)$.  Let $p_s(\tau),p_s^h(\tau)$ be
the trapping time distributions both at zero field and after applying
a field $h$ at time $s$ respectively. Eq.\eq{g12} holds whenever
$p_s^h(\tau)=p_s(\tau)+{\cal O}(h^2)$~\footnote{Similarly one could
say that the average trapping time is not modified to linear order in
$h$}.  In other words, up to linear order in $h$ the sole effect of
the field is to modify the density of configurations as indicated
in\eq{g6}.

The physical significance of these results can now be appreciated at
its full extent. In the simplest partial equilibration scenario the
effective temperature derived from the microcanonical relation
\eq{g13} is independent on the type of observable and coincides with
that obtained from the FT\eq{n3} and the
FDR\eq{ad2b}. For the last equality, the trapping time distribution
is required to remain unchanged at linear order with the intensity of
the field $h$. In this case,
\be
T_{\rm eff}^{{\rm FDR}}(s)=T_{\rm eff}^{\rm FT}(s)=T_{\rm eff}^{{\rm MR}}(s)
\label{g14}
\ee
This equality is remarkable as it shows that the effective temperature
can be directly obtained from the fluctuation theorem\eq{n3} without
the necessity of introducing a perturbing field and measuring
correlations and responses~\footnote{From a different perspective, a
result where a perturbing field is not required to measure the FDR has
been recently proposed~\cite{Chatelain03,Ricci03}.}. The first
equality in\eq{g14} has been numerically verified in a given example
of spin-glass model~\cite{CriRit03b}. In what follows we exemplify these results for a simple
solvable case.

\section{The oscillator model (OM)}
\label{om}

Exactly solvable oscillator models (OMs) \cite{BonPadRit97}, unrealistic as they look,
provide a simple physical basis to describe glassy dynamics. A
parallel can be established between oscillator models for glassy
dynamics and the original urn models by the Ehrenfest's. The former can
enlighten our comprehension of the essential features of glassy
dynamics, in the same way urn models have provided a ground basis to
understand key concepts of equilibrium thermodynamics such as the
Boltzmann entropy.  The OM is an example where the partial
equilibration scenario holds for the dynamics at $T=0$. It provides an
excellent framework to verify the statements of previous sections.

\subsection{Energy relaxation}
\label{reminder}

The OM \cite{BonPadRit97} is described by a set of continuous variables $x_i$
and an energy function,
\be
E=\frac{K}{2}\sum_{i=1}^N x_i^2
\label{b1}
\ee
where $K$ is the spring constant and $N$ is the total number of
oscillators. Oscillators are non-interacting and therefore the model
has trivial static properties, the total energy $E=Nk_BT/2$ according
to the equipartition law. We consider a cooperative dynamics where
oscillators are updated according to the rule $x_i\to
x_i+\frac{r_i}{\sqrt{N}}$, the $r_i$ being uncorrelated Gaussian
variables with $\overline{r_i}=0$ and variance $\overline{r_i
r_j}=\Delta^2\delta_{ij}$. The updating of all oscillators is carried
out in parallel and the moves are accepted according to the Metropolis
rule. We will focus our analysis on the dynamics at $T=0$ where only
updates that decrease the total energy are accepted. This leads to
slow dynamics as most of the proposed moves tend to increase the
energy and only few of them are accepted (i.e. the acceptance rate is
quickly decreasing with time).  Dynamics in the OM is ergodic if
confined to a constant energy surface (see
Fig.~\ref{ergo}). Therefore, dynamics does not become arrested at
$T=0$ as no metastable configurations (except the ground state) exist
at $T=0$.

Dynamical properties in the OM are derived from the distribution of
attempted energy changes $P(\Delta E)$. There are several ways to
compute this distribution \cite{pde}, the simplest one derives from the
Gaussian character of such distribution. The change in
energy of an elementary move is given by
\be
\Delta E=\frac{K}{\sqrt{N}}\sum_{i}\,x_ir_i+\frac{K}{2N}\sum_ir_i^2     .
\label{b2}
\ee
From the Gaussian character the of $r_i$, it follows that $\Delta E$ has a 
Gaussian distribution whose mean and variance are given by,
\be
M_{\Delta
E}=\overline{\Delta E}=K\Delta^2/2,~~~~~~\sigma_{\Delta
E}=\overline{(\Delta E)^2}-(\overline{\Delta E})^2=2K\Delta^2\frac{E}{N}=2K\Delta^2 e~~~
\label{b3}
\ee
which yields \cite{BonPadRit97},
\be P(\Delta E)=\bigl(4\pi eK\Delta^2\bigr)^{-\frac{1}{2}}
      \exp\left[-\frac{(\Delta E-\frac{K\Delta^2}{2})^2}{4eK\Delta^2}
           \right]
\label{b4}
\ee
where $e=E/N$ is the energy per oscillator. At $T=0$, the dynamical evolution of
the energy $e$ and the acceptance rate $a(e)$ (the
fraction of accepted moves) are given
by the following closed equations,
\be
\frac{\partial e}{\partial
t}=\int_{-\infty}^{0}xP(x)\,dx~~~~~;~~~~a(e)=\int_{-\infty}^0dx P(x)=\frac{1}{2}{\rm Erfc}(\sqrt{\frac{K\Delta^2}{16e}})
\label{b5}
\ee
with ${\rm Erfc}(x)=(2/\sqrt{\pi})\int_x^{\infty}du\exp(-u^2)$ the
complementary error function.
As the energy $e$ decreases the variance of the distribution\eq{b3}
decreases. Accordingly, the acceptance rate also decreases. The dynamical evolution
of the energy and acceptance can be solved in the long-time limit, $E(t)\sim
1/\log(t), a(t)\sim 1/(t\log(t))$. 

\subsection{Effective temperature}
\label{teffom}

Correlations and responses have been computed for the magnetization
$M=\sum_i x_i$ \cite{BonPadRit97}. In the rest of the paper we will consider $A=M$ as the
observable under quest. $M$ is a neutral observable as can be verified
by solving the dynamical equation for the magnetization.  It relaxes
exponentially fast to zero which is the equilibrium value of the
magnetization at all temperatures.  The autocorrelation function
$C_M(t,s)=(1/N)\sum_i x_i(t)x_i(s)$, the corresponding response
function $R_M(t,s)$\eq{ad2} and the susceptibility
$\chi_M(t,s)=\int_{s}^tR_M(t,t')dt'$, do not have stationary part but
only aging part and show simple aging with $\tau_{\rm
decorr}(s)\propto s$.  Correlations and responses define an effective
temperature $T_{\rm eff}(t,s)$ through the relation,
\be
T_{\rm eff}^{\rm FDR}(t,s)=\frac{\frac{\partial C_M(t,s)}{\partial s}}{R_M(t,s)}
\label{b6}
\ee
which is the temperature $T$ of the bath in equilibrium. At $T=0$ the
response is finite in the OM due to the shift of energy levels
described in Sec.~\ref{teff}. A simple expression can be derived in that case,
\be
T_{\rm eff}^{\rm FDR}(t,s)\equiv T_{\rm eff}^{\rm FDR}(s)=2e(s)+\frac{1}{f(s)}\frac{\partial
e(s)}{\partial s}
\label{b7}
\ee
where $f(s)$ is a given function \cite{BonPadRit97} which
asymptotically decays as $1/s$. Two remarkable facts emerge
from\eq{b7}: 1) $T_{\rm eff}(t,s)$ only depends on the lowest time
$s$, therefore characterizes the aging state of the system at time
$s$; 2) The second term in the r.h.s of\eq{b7} is sub-dominant respect
to the first term leading to $T_{\rm eff}(s)\to 2e(s)$, i.e. the
equipartition law is asymptotically satisfied in the aging regime.

\subsection{The fluctuation theorem}
\label{ftom}
As the energy decays logarithmically and $\tau_{\rm decorr}(s)\propto
s$,\eq{ad4} is satisfied and relaxation is adiabatic. Moreover,
dynamics in this model is ergodic if constrained to the constant
energy surface. Under these conditions, only spontaneous relaxation
occurs (Sec.~\ref{ss}) and partial equilibration holds (Sec.~\ref{partial}). 
We can verify the validity of the FT by substituting\eq{b4} in\eq{n2},
\be \frac{P(\Delta E)}{P(-\Delta E)}=\exp\Bigl(\frac{\Delta E}{T_{\rm
 eff}^{\rm FT}}\Bigr)=\exp\Bigl(\frac{\Delta E}{2e}\Bigr)
\label{f1}
\ee
giving $T_{\rm eff}^{\rm FT}=2e$. This result coincides with the
asymptotic value $T_{\rm eff}^{\rm FDR}$ previously obtained~\eq{b7}.

\subsection{Trapping time distribution}
\label{trapping}
Here we compute the trapping time distributions without and in a field
$p_s(\tau),p_s^h(\tau)$. In the presence of a field the energy of the OM is given by,
\be
E=Ne_T=\frac{K}{2}\sum_{i=1}^N x_i^2-h\sum_{i=1}^N x_i=N(e-hm)
\label{c1}
\ee where $e,m$ denote $(K/2)\overline{x^2}$ and $\overline{x}$ (i.e.
the energy and magnetization per oscillator respectively), the total
energy per oscillator being $e_T=e-hm$.  Eq.\eq{c1} can be written as
$E=\frac{K}{2}\sum_{i=1}^N (x_i-h/K)^2-Nh^2/(2K)$. The updating rule
for the shifted variables $y_i=x_i-h/K$ remains
unchanged. Consequently, the evolution equation for the quantity
$e'=(K/2)\overline{y^2}=e-hm+h^2/(2K)=e_T+h^2/(2K)$ is identical to
that obtained for $e$ at $h=0$. As the time evolution of $e'$ does not
depend on $h$ this implies that $e_T$ gets corrections that are even
powers of $h$. In a field,\eq{b5} holds by replacing $e$ with
$e'$. Therefore, the acceptance rate is not modified at linear order
in $h$. In general, the trapping time probability distribution is
given by $p_s^h(\tau)=a^h(e')(1-a^h(e'))^{\tau-1}$ where $\tau$ is the
finite number of Monte Carlo steps (i.e. does not scale with $N$) and
$a^h(e')$ is the acceptance rate in a field. The same expression is
valid for $p_s(\tau)$ putting $h=0$. This immediately proves that the
distribution of trapping times remains unchanged at linear order in
$h$. In particular, the average trapping time is finite and given by
the inverse of the acceptance rate,
\be
\overline{\tau^h (e')}=\frac{\tau_0}{\int_{-\infty}^0P_h(\Delta E)d(\Delta E)}=\frac{\tau_0}{a^h(e')}
\label{c2}
\ee
where $\tau_0$ is a microscopic time. 
\subsection{Microcanonical rates for  the magnetization}

Let us consider the joint probability $P^h(\Delta E,\Delta M)$ of
having a change in the total energy $\Delta E$ and magnetization
$\Delta M$ in the presence of an external field $h$ ($E$ is given in
\eq{c1} and includes the Zeeman term). At $T=0$, the probability
$a^h_E$ that an attempted change $\Delta M$ is accepted is given by,
\be W^h_E(\Delta M)=\frac{1}{\tau_0}\int_{-\infty}^0P^h(\Delta
E,\Delta M)d(\Delta E)=\frac{1}{\tau_0}P^0(\Delta
M)\int_{-\infty}^0P^h_c(\Delta E|\Delta M)d(\Delta E)~~~.
\label{d1}
\ee
where $P^h(\Delta E,\Delta
M)=P^h_c(\Delta E|\Delta M)P^0(\Delta M)$, the subscript $c$ standing
for conditional probability~\footnote{We are following the notation of Sec.~\ref{teff}
where $W_E(\Delta M)$ stands for the rate at zero field.}. Again, it is easy to show that these
distributions are all Gaussian. Straightforward calculations give,
\bea
P^0(\Delta M)=(2\pi\Delta^2)^{-\frac{1}{2}}\exp\Bigl[ -\frac{(\Delta
M)^2}{2\Delta^2} \Bigr]\label{d3a}\\
P^h_c(\Delta E|\Delta
M)=(2\pi\alpha_h^2)^{-\frac{1}{2}}\exp\Bigl[-\frac{(\Delta
E-\frac{K\Delta^2}{2}-b_h\Delta M)^2}{2c_h^2\sigma_h^2} \Bigr]\label{d3b}
\eea
with the definitions $\alpha_h=c_h\sigma_h$, $b_h=g_h\sigma_h^2$,
$\sigma_h^2=2Ke+h^2-2Khm$, $c_h^2=\Delta^2(1-\frac{(Km-h)^2}{\sigma_h^2})$,
$g_h=(Km-h)/(2Ke+h^2-2Khm)$. In the linear response regime we
obtain, $g_h=\frac{m}{2e}+(m^2/(2e^2)-1/(2Ke))h + {\cal O}(h^2)$,
$c_h^2=c_0^2(1+mh/e+{\cal O}(h^2))$, $c_0^2=\Delta^2(1-m^2 K/(2e))$,
$\sigma_h^2=2Ke(1-hm/e+{\cal O}(h^2))$,$b_h=mK-h+{\cal
O}(h^2)$.  Inserting\eqq{d3a}{d3b} in\eq{d1} gives,
\bea W^h_E(\Delta M)=W_E(\Delta M)\exp\Bigl[\frac{(\Delta M)
h}{4e-2m^2K}(1+\frac{2m\Delta M}{\Delta^2})\Bigr]\label{d4a}\\
W_E(\Delta M)=\frac{1}{\tau_0}\frac{P^0(\Delta M)}{2}{\rm Erfc}\Bigl[\frac{(k\Delta^2)/2+b_0 (\Delta
M)}{2^{1/2}\alpha_0}\Bigr]
\label{d4b}
\eea
where $P^0(\Delta M)$ is given by\eq{d3a}. As $m$ is a neutral
observable that relaxes fast to $0$ we can replace $m=0$ everywhere in
all previous expressions. Up to linear order in $h$ the rates $W_E^h$ are given by,
\be
W^h_E(\Delta M)=W_E(\Delta M)\exp\Bigl(\frac{\Delta M
h}{4e}\Bigr)~~~~~~;~~~~~~~~W_E(\Delta M)= P^0(\Delta M) a(e)
\label{d5}
\ee
where $a(e)$ is given in\eq{b5}. We remark the following results: 1) the perturbed rates are multiplicative~\footnote{The
form of the perturbed rates\eq{d5} is multiplicative. This assumption
was made in a given class of trap models at the level of
configurations and shown to lead to the existence of effective
temperatures~\cite{Ritort03}. Although the multiplicative property may hold at
the coarse-grained level of observable values, it was shown to be
far-fetched at the level of configurations~\cite{Sollich03}. Eq.\eq{d5} shows
that rates are to be considered multiplicative only at the level of
observable values.}, 2)\eq{g4} holds as $W_E(\Delta M)=W_E(-\Delta M)$ and 3) relation\eq{g10} is satisfied as well,
\be
\frac{W^h_E(\Delta M)}{W^h_E(-\Delta M)}=\frac{W_E(\Delta M)}{W_E(-\Delta M)}\exp\Bigl[\frac{(\Delta M)  h}{2T_{\rm eff}^{\rm MR}(e)}\Bigr]
\label{d6}
\ee
so the effective temperature here obtained $T_{\rm eff}^{\rm MR}(e)=2e$ again
coincides with that derived from the FDR\eq{b7} and the FT\eq{f1}.

\section{Conclusions}

A statistical interpretation of a non-equilibrium or effective
temperature in aging systems, rather than a thermometric one, might be
possible. Recent numerical simulations of a disordered model
\cite{CriRit03b} have shown that the effective temperature, as
measured from FDT violations, originate from the existence of a {\em
spontaneous} relaxational process describing heat exchange
low-frequency events.  Superimposed to it there is a {\em stimulated}
process that is characterized by the temperature of the bath and
responsible of most of the heat exchange observable events between
system and bath.  Although the timescale of the spontaneous process is
related to the temperature of the bath, its statistical description
(e.g. the specific form of the corresponding FT) is related to other
properties of the aging state such as the energy content. The
concurrence of these two processes in the overall relaxation is
related to the intermittent phenomenon recently observed in various
experiments~\cite{Cil03,Cip03}. Moreover, the width of the exponential
tail associated to the spontaneous process is predicted to be
proportional to the effective temperature.

We have investigated a partial equilibration scenario where only the
spontaneous process occurs (the temperature of the bath is set to zero)
and dynamics is ergodic when constrained to a given energy surface.  In
this case three effective temperatures can be defined: a) from the
fluctuation-dissipation ratio\eq{ad2b}; b) from the fluctuation theorem
for the statistical description of the spontaneous component\eq{n1}; c)
from microcanonical relations relating observable changes\eq{g13}. All
three are shown to be identical\eq{g14}. Explicit calculations in the OM
demonstrate the validity of these statements. 

Several open questions and directions of research along the present
ideas appear worthwhile. It would be useful to go beyond the qualitative
level of demonstrations in the present paper over more founded
mathematical proofs of all these results. Attempts trying to establish
the origin of effective temperatures using master equation formalisms
have already appeared in the literature~\cite{Chatelain03,Diezemann03}
and we foresee more work in the near future. We should also mention  
the close similarity between the present approach and that by Edwards
for granular matter~\cite{Edwards89}.
Would be very interesting
to investigate other classes of models with simple equilibrium
properties, such as kinetically constrained models~\cite{RitSol03},
where the existence of non-trivial effective temperatures is still under
debate. In these models it is possible to identify correlated motion of
particles~\cite{GarCha03} that underpin a spatially heterogeneous
dynamics~\cite{BerGar03}.  Another category of interesting problems to
explore are systems in steady states where FDT violations have been
studied~\cite{SanRegRub01} and where non-Gaussian effects,
similar to those described here, have been identified as
well~\cite{ZohCoh03}.


\begin{thebibliography}{99}
\bibitem{pde} A detailed exposition can be found in Sec.~6.5.1 of \protect\cite{CriRit03}.
\end{thebibliography}
\end{document}